# FREE ELECTRON LASER SEEDED BY BETATRON RADIATION

A. Ghigo, M. Galletti, V. Shpakov, INFN-LNF, Frascati, Italy
A. Curcio, CLPU, Villamayor, Salamanca, Spain
V. Petrillo[1], Milan University, Milano, Italy
[1]also at INFN-Mi, Milano, Italy

*Abstract*

In this paper the use of betatron radiation as a seed for the Free Electron Laser (FEL) is presented. The scheme shown can be adopted from all FEL driven by plasma accelerated electron beams via Particle or Laser Wake Field Acceleration. Intense radiation in the region of X ray characterized by a broad spectrum, the betatron radiation, is indeed produced in the plasma acceleration process from the electron passing through the ionized gas. It is proposed to use this radiation, suitably selected in wavelength and properly synchronized, to stimulate the emission of the Free Electron Laser.

## INTRODUCTION

The possibility of using a plasma accelerated electron beam to generate Free Electron Laser (FEL) radiation has recently been proven [1]. The European EuPRAXIA project aims to develop FEL facilities using laser wake field acceleration and particle wake field acceleration (PWFA) techniques. In particular, in the INFN Frascati Laboratories, the headquarters of the EuPRAXIA project, an infrastructure will be built that will use the PWFA to generate FEL radiation in X-rays region [2]. The electron beam produced by a low emittance injector is accelerated by an X-band linac and a plasma acceleration section. Due to the intense transverse forces generated in the plasma wave, the electrons of the bunch oscillate emitting the so called betatron radiation. The radiation is emitted in a wide bandwidth in the X ray region and the basic idea is to select a narrow portion of the betatron radiation spectrum with a monochromator and to send this radiation superimposed on the same electrons beam that generated it, towards the magnetic undulator, as shown in Fig.1[3]. The betatron radiation acts as seed of the Free Electron Laser emission: in EuPRAXIA, if the selected photon energy is matched with undulator fundamental wavelength (4nm), the seeding scheme enhances the number of photons per pulse and improves the pulse-to-pulse temporal stability as happens in the self-seeding scheme [4]. To extend the FEL emission spectrum towards high frequencies, the gap of the undulator can be opened. However, in these conditions, the SASE gain is no longer sufficient to saturate with the parameters of the beam expected at the accelerator output and with the nominal undulator length. It can be shown that by injecting the selected betatron radiation at the highest frequency that the most open undulator would have, it is possible to considerably extend the range of frequencies in which the FEL can saturate.

## EXPERIMENTAL SET-UP

In the particle wake field plasma acceleration, a high charge electron bunch, the *driver* bunch, generates the plasma shape, losing energy, while a following low charge bunch, the *witness* bunch, is accelerated. Before entering in the FEL undulator, at the plasma chamber exit, the *driver* and *witness* bunches are dispersed, due to the different energy, in a dispersive section composed by four dipole magnets in a chicane configuration. The deviation angle in the first chicane dipole is large enough to separate the witness from the driver bunch. A collimator, or beam scraper, can be placed to stop the driver bunch. The betatron radiation propagates straight into the vacuum chamber and, as soon as the electron beam is deflected from the straight path, the first optical element of the monochromator is installed.
The betatron radiation is selected in bandwidth and reflected back in the direction of the undulator overlapping the electrons that are leaving the chicane, in the first part of the undulator. Because of the very short electron bunch a perfect synchronization at the entrance of the undulator, at level of tens of femtosecond, is needed. The electron and photon beams started automatically synchronized because generated from the same electron beam. The trajectory length of the photons in the monochromator must compensate the path of the electrons that pass through the magnetic chicane and the delay of the electron, that travel with relativistic factor γ=2000, respect to the photon arrival time.

## BETATRON RADIATION

Betatron radiation is the radiation emitted by electrons accelerated in plasma channels. Betatron radiation is emitted forward, and it is due to the betatron oscillations driven by the focusing fields inside the plasma bucket. Due to the very short scale of the betatron oscillations period (typically from mm down to microns, depending on the background electron plasma density), the typical energy of the photons emitted via betatron radiation falls in the X-rays.

*Theoretical Introduction*

If the scale of the electron energy gain is much longer than the betatron period, it can be assumed that the acceleration occurs adiabatically compared to the betatron dynamics. This allows using the following formula for the energy irradiated *I* by a single electron per unit photon energy *E* and

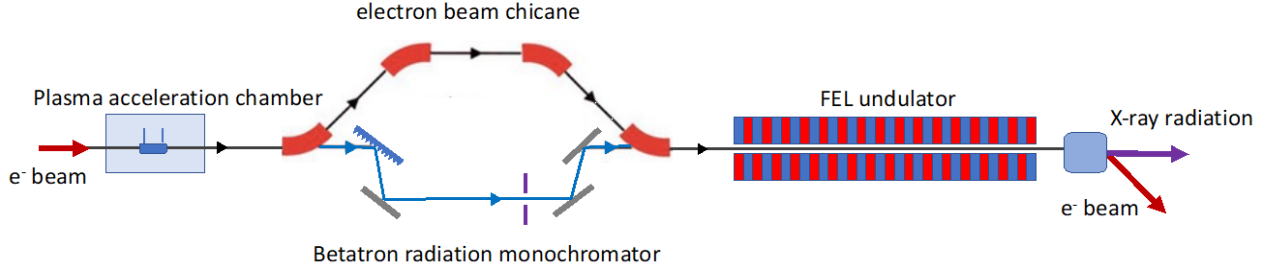

Figure 1: Betatron radiation seeding FEL scheme.

solid angle of observation $\Omega$:

$$\left.\frac{d^2I}{dEd\Omega}\right|_{single} = \int \frac{dz}{L_{acc}} \sum_n \frac{\alpha N_\beta \gamma^2(z) E R_n}{2\hbar\omega_\beta}[C_r^2 + C_z^2\theta^2 - 2C_rC_z\theta\cos\varphi] \quad (1)$$

where $\alpha$ is the fine structure constant, $N_\beta = \omega_\beta L_{acc}/c$ is the number of betatron oscillations over the acceleration length $L_{acc}$ ($z$ is the acceleration axis here), $c$ is the speed of light in vacuum, $\gamma(z)$ is the Lorentz factor of the emitting electron, $\omega_\beta = \omega_p/\sqrt{2\gamma(z)}$ is the betatron frequency related to the plasma frequency $\omega_p = \sqrt{n_e e^2/\varepsilon_0 m_e}$, where $n_e$ is the background electron plasma density, $m_e$ is the electron mass, $e$ is the elementary charge and $\varepsilon_0$ is the vacuum dielectric constant. The summation in Eq. (1) runs over $n$, the harmonic numbers of the resonance function $R_n$. Indeed, the plasma channel acts similarly as a wiggler device, therefore lines emission should be in principle expected. In reality, we shall see that being the harmonic frequency related to the betatron amplitude, the finite size of the beam is responsible for spectral broadening. Moreover, in Eq. (1) we have defined:

$$C_r = r_\beta k_\beta \sum_m J_m(\rho_z)[J_{n+2m-1}(\rho_r) + J_{n+2m+1}(\rho_r)] \quad (2)$$

$$C_z = \sum_m 2J_m(\rho_z)J_{n+2m}(\rho_r) \quad (3)$$

$$\rho_z = \frac{E r_\beta^2 k_\beta^2}{8\hbar\omega_\beta} \quad (4)$$

$$\rho_r = \frac{E r_\beta k_\beta \theta \cos\varphi}{\hbar\omega_\beta} \quad (5)$$

$$R_n = sinc^2\left[\frac{\pi N_\beta}{2\gamma^2(z)\hbar\omega_\beta}\left(E - \frac{2n\gamma^2(z)\hbar\omega_\beta}{1+\frac{K_\beta^2}{2}+\gamma^2(z)\theta^2}\right)\right] \quad (6)$$

where $r_\beta$ is the betatron amplitude, $K_\beta = \gamma(z)r_\beta k_\beta$ is the undulator parameter of the plasma wiggler and $k_\beta = \omega_\beta/c$. Concluding the present section, we introduce the final formula for the calculation of the spectral-angular distribution of the betatron radiation that includes the distribution of the betatron oscillation amplitudes. Indeed, each electron oscillates with a different amplitude $r_\beta$, according to the position within the beam that it had at the entrance of the plasma channel. Given a radial profile of the beam $P(r_\beta)$ (we assume radial symmetry here), the betatron radiation formula becomes:

$$\left.\frac{d^2I}{dEd\Omega}\right|_{beam} = \frac{Q}{e}\int_0^\infty dr_\beta\, r_\beta\, P(r_\beta)\left.\frac{d^2I}{dEd\Omega}\right|_{single} \quad (7)$$

## Betatron Radiation from the Witness Bunch

The target parameters of the witness bunch in Eupraxia are resumed in the table below.

Table 1: Witness Bunch Electron Beam Parameters

|  | In | Out |
| --- | --- | --- |
| Charge | 30 pC | 30 pC |
| Beam size (transverse, rms) | 2 µm | 2 µm |
| Beam size (longitudinal, rms) | 7 µm | 7 µm |
| Normalized rms emittance | 0.6 mm mrad | 0.6 mm mrad |
| Relative energy spread, rms) | 0.05 % | 0.05 % |
| Peak current | 1.8 kA | 1.8 kA |
| Beam energy (mean) | 500 MeV | 1 GeV |

For the simulation of the radiated betatron spectrum we have assumed a linear energy gain, i.e. $\gamma(z)$ is a linear function of $z$, with initial value $\sim 1000$ at $z=0$ and final value $\sim 2000$ at $z=0.4\, m = L_{acc}$. Furthermore, we consider a gaussian beam and the background electron plasma density is $n_e = 3 \times 10^{16} cm^{-3}$. The result of the simulation based on Eq. (7) and on the data reported in table above is shown in Fig. 1.

From Fig. 2 it is possible to infer the number of emitted photons, after integration over $E$. In particular, the number of photons emitted at *620 eV (2 nm)*, within a bandwidth of *10 %*, is $8.9 \times 10^6$, while the number of photons emitted at *310 eV (4 nm)* within a bandwidth of *10 %*, is $1.4 \times 10^7$. Reducing the bandwidth to *1 %*, we get $1.4 \times 10^6$ at *4 nm* and $8.9 \times 10^5$ at *2 nm*.

## SEEDED FEL SIMULATIONS

The betatron radiation has been used as seed for the FEL emission generated by the witness electron beam. The undulator is the EuPRAXIA high-energy line AQUA [5], with

10 modules with period $\lambda_w$=1.8 cm for a total length of about 25 m. The initial longitudinal power distribution of the seed has been prepared with random spikes and random phase structure so as to mimic the incoherent structure of the betatron pulse. The witness beam has been matched to the undulator with transverse dimensions $\sigma_x$=69 μm and $\sigma_y$=44 μm. The SASE simulation, made with GENESIS 1.3 [6] at 4 nm (Fig. 3, blue curve) shows that the radiation in 25 m is still in the exponential stage, achieving at the undulator end 5.3 μJ of energy, corresponding to $10^{11}$ photons/shot. Starting with the seed, instead, allows the radiation to reach saturation within 20-22 m, arriving to an energy of 20 μJ, corresponding to $4 \cdot 10^{11}$ photons/shot (red curves). The stability of the pulse is moreover increased.

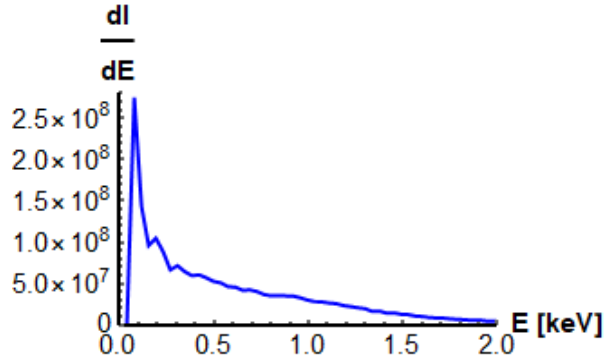

Figure 2: Simulated betatron radiation spectrum emitted by the Eupraxia witness bunch.

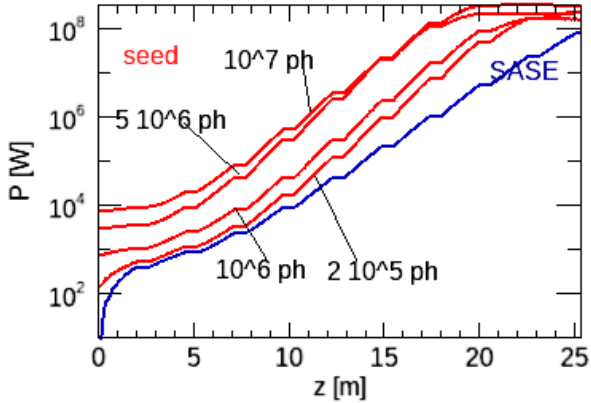

Figure 3: Growth of the FEL radiation at 4 nm along z(m) for the SASE case (blue curve) and various seed values (red curves)

At 3.2 nm (see Fig. 4), the use of the betatron radiation as a seed appears even more advantageous, provided to seed the FEL with at least $2/3 \cdot 10^6$ photons. In this case, in fact, the SASE provides $3 \cdot 10^9$ photons, vs $10^{11}$ of the seeded operation. Table 2 summarizes the data.

## CONCLUSION

The betatron radiation, produced during plasma acceleration by transversely oscillating electrons, can be efficiently used as seeding for FEL.

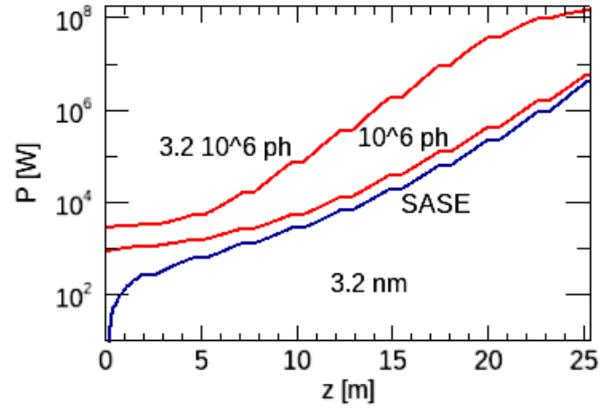

Figure 4: Growth of the FEL radiation at 3.2 nm along z(m) for the SASE case (blue curve) and various seed values (red curves)

Table 2: Summary of FEL Radiation Photon Number and Bandwidth at 4 and 3.2 nm for Different Seed Energies.

| $\lambda_{rad}$ | $N_{seed}$ ph. | $N_{rad}$ ph. | BW |
|---|---|---|---|
| 4 nm | 0 | $10^{11}$ | 0.08% |
| 4 nm | $10^6$ | $3 \cdot 10^{11}$ | 0.12 % |
| 4 nm | $5 \cdot 10^6$ | $4 \cdot 10^{11}$ | 0.12% |
| 3.2 nm | 0 | $3 \cdot 10^9$ | 0.1% |
| 3.2 nm | $10^6$ | $5 \cdot 10^9$ | 0.09% |
| 3.2 nm | $2 \cdot 10^6$ | $10^{11}$ | 0.09% |

The scheme has been proved to be adoptable, from all FEL driven by plasma accelerated electron beams via Particle or Laser Wake Field Acceleration, suitably selecting the suitable wavelength with a proper synchronization, to stimulate the emission of the Free Electron Laser. Among next steps, we will consider the betatron radiation produced by the driver bunch which carries much higher charge and a different spectrum due to the energy. Moreover, the betatron field propagation with the self-consistent phase input for the FEL simulations will be optimized. A more accurate simulation campaign accounting for 3D effects in betatron radiation emission, especially the impact on the final photon number, is envisioned. Finally, having the betatron source optimized, simulations on the betatron radiation stimulating the emission on the higher harmonics will be carried out.